\documentclass{pasj02}
\usepackage[switch,mathlines]{lineno}

\Received{2024/07/13}
\Accepted{$\langle$acception date$\rangle$}
\Published{$\langle$publication date$\rangle$}
\begin{document}

\title{A search for water vapor plumes on Europa by spatially-resolved spectroscopic observation using Subaru/IRCS}
\author{Jun \textsc{Kimura}\altaffilmark{1,}$^{*}$, Taro \textsc{Matsuo}\altaffilmark{2}, Hitomi \textsc{Kobayashi}\altaffilmark{3}, Yuji \textsc{Ikeda}\altaffilmark{4,5}, Kazuo \textsc{Yoshioka}\altaffilmark{6}, Seiko \textsc{Takagi}\altaffilmark{7} and Shigeru \textsc{Ida}\altaffilmark{8}}%
\altaffiltext{1}{Department of Earth and Space Science, Graduate School of Science, Osaka University, 1-1 Machikaneyama-cho, Toyonaka, Osaka 560-0043, Japan}
\altaffiltext{2}{Department of Particle and Astrophysics, Graduate School of Science, Nagoya University Furocho, Chikusa-ku, Nagoya, Aichi 466-8601, Japan}
\altaffiltext{3}{Photocross Co., Ltd., 819-1, Iwakura-Chuzaichi-cho, Sakyo-ku, Kyoto 606-0021, Japan}
\altaffiltext{4}{Photocoding, 460-102 Iwakura-Nakamachi, Sakyo-ku, Kyoto, 606-0025, Japan}\altaffiltext{5}{Laboratory of Infrared High-resolution Spectroscopy (LiH), Koyama Astronomical Observatory, Kyoto Sangyo University, Motoyama, Kamigamo, Kita-ku, Kyoto 603-8555, Japan}
\altaffiltext{6}{Graduate School of Frontier Sciences, The University of Tokyo, 5-1-5, Kashiwa-no-ha, Kashiwa 277-8561, Japan}
\altaffiltext{7}{Department of Earth and Planetary Sciences, Faculty of Science, Hokkaido University, Kita 10, Nishi 8, Kita-ku, Sapporo 060-0810, Japan}
\altaffiltext{8}{Earth-Life Science Institute, Tokyo Institute of Technology, 2-12-1-IE-1 Ookayama, Meguro-ku, Tokyo 152-8550, Japan}

\email{junkim@ess.sci.osaka-u.ac.jp}

\KeyWords{planets and satellites: atmospheres ---  planets and satellites: individual (Europa) ---  techniques: spectroscopic}

\maketitle

\begin{abstract}
We present near-infrared high-dispersion spectroscopic observations of Europa using the Infrared Camera and Spectrograph (IRCS) onboard the Subaru Telescope, seeking direct evidence of water plumes on Europa and exploring spatial variations in plume activity.
Using high spectral/spatial resolution and sensitivity of Subaru/IRCS, our observations have enabled a spatially resolved search for water plumes on Europa.

Within our detection limits and time of observation, we found no evidence for the presence of water emission. 
For a rotation temperature of 50 K, we derived an upper limit on the H$_{2}$O abundance of 9.46×10$^{19}$ – 5.92×10$^{20}$ m$^{-2}$ in each divided slit area and 4.61×10$^{19}$ m$^{-2}$ in the entire area covered by the slit.
This upper limit lies below the inferred water abundance from previous UV observations by the Hubble Space Telescope (HST), while being less sensitive by a factor of three compared to the Keck telescope and by one order of magnitude or more than the James Webb Space Telescope (JWST) observations.
Our results align with previous studies and demonstrate that using Subaru/IRCS is an effective strategy for searching for water plumes on Europa with high spatial resolution.
Continued observations across different surface areas and orbital phases are essential to fully characterize Europa's plume activity and complement upcoming space missions.
\end{abstract}
%
%\pagewiselinenumbers
%
\section{Introduction}
Jovian moon Europa stands out as a primary target for astrobiological exploration in the solar system.
The Galileo spacecraft's encounter with Europa revealed magnetic field fluctuations, indicating a responsive behavior of Europa's interior conductors to changes in the Jovian magnetic fields \citep{kivelson00}.
This suggests the presence of a sub-surface global ocean containing electrolytes beneath the solid ice shell \citep{khurana09}.
Gravitational measurements, geomorphological interpretations  and numerical calculations suggest to host an ocean which has several tens of kilometers thickness, beneath the solid ice shell with few to few tens of kilometers thickness \citep{anderson98,pappalardo98,schenk02,kimura24}.
The ocean is likely to be enriched in the materials and chemical energy necessary to sustain life \citep{vance23}, thus its chemical characterization will help to understand the potential habitability in Europa.
A report of a possible water eruption from UV observations with the Hubble Space Telescope \citep{roth14b}, when Europa was near apojove (true anomaly $\sim$182-218$^{\circ}$), suggests that the possibility that Europa's oceanic material could be ejected into space, providing direct access to the material without the need for drilling through the km-thick ice shell.
\citet{roth14b} estimated total content of 2\,$\times$\,10$^{31}$ water molecules, average column density of 1.5\,$\times$\,10$^{20}$\,molecules/m$^{-2}$ in the southern hemisphere and production rate of 6.6\,$\times$\,10$^{28}$ molecules\,s$^{-1}$ assuming a residence time in the plume of $\sim$\,10$^{3}$ seconds.
Additionally, magnetic--field and plasmawave observations of the Galileo spacecraft during a close flyby of Europa have been interpreted as being caused by a water plume \citep{jia18}.
On the other hand, Europa's atmosphere is also fed with materials stripped from its icy surface due to energy radiation from Jovian plasma and thermal desorption by solar radiation, and meteoritic impacts.
As a result, water, oxygen, and hydrogen molecules are estimated to be populated at a rate of about 10$^{26-27}$ molecules\,s$^{-1}$ \citep{vorburger18}.
Therefore, estimated abundances of water vapor from the plume could be exceed from exogenic effects by factors 10s--100s.
However, such plume activity has not been confirmed by subsequent observations despite several attempts \citep{roth14a}.
The reliability of direct imaging observations by the HST during Europa's transit in front of Jupiter (at sub-observer longitude of $\sim$182-186$^{\circ}$$\,\pm\,$40$^{\circ}$) for true anomaly of 260$^{\circ}$, 297$^{\circ}$ and 310$^{\circ}$, suggesting a total of 1.3$\times$10$^{32}$ molecules and an column density of 1.5\,$\times$\,10$^{20}$\,molecules/m$^{-2}$ \citep{sparks16}, has been questioned, attributing potential statistical errors \citep{giono20}. 
Subsequent attempt with different approach has failed to confirm these findings \citep{sparks19}.
Infrared spectroscopy by the Keck telescope identified one tentative detection out of 17 dates with a total of (7.0\,$\pm$\,2.2)$\times$10$^{31}$ water molecules and a column density of (1.4\,$\pm$\,0.4)\,$\times$\,10$^{19}$\,molecules/m$^{2}$, at sub-observer longitude of $\sim$134-150$^{\circ}$ and Europa's true anomaly of 159-176$^{\circ}$ \citep{paganini20}.
The James Webb Space Telescope observation of Europa's leading hemisphere on one date yielded no detection of water vapor, resulting in a 3$\sigma$ upper limit of 2.71\,$\times$\,10$^{18}$\,m$^{-2}$, which is derived from the upper limit of 3.5\,$\times$\,10$^{31}$ water molecules and the integrated area of 1.3-arc sec-diameter (Europa's disk is 1.0-arc sec-diameter) \citep{villanueva23}.
These results strengthen the argument that potential plumes on Europa are primarily confined to specific areas, occur infrequently, exhibit low intensity, or contain minimal volatile components.
Even if present, details regarding its temporal and spatial variations remain unknown.
Such uncertainty arises from the lack of spatial resolution in previous spectral observations.
Therefore, reducing this uncertainty is crucial for better understanding the characteristics of Europa's plume activity.

We conducted spatially-resolved high-dispersion near-infrared spectroscopic observations using the Subaru Telescope/IRCS to search for the direct evidence of water arising from Europa's plumes and, if present, to explore spatial variations in plume activity.
Section 2 describes the observations with Subaru/IRCS, methodology, and data reduction method.
In Section 3, we show our results based on the near-infrared spectroscopic observations of Europa.
Using our observational data, we estimate the water abundance for each observed area on Europa.
In Section 4, we discuss comparisons of our results to previous observations.
Finally, we summarize our results and suggest implications for plume activity on Europa in Section 5.

%%%%%%%%%%%%%%%%%%%%%%%%%%%%%%%%%%%%%%%%%%%%%%%%%%%%%%%%%%%%%%
\section{Observations and Data Analysis}
\subsection{Observations using Subaru/IRCS}\label{section2.1}
\begin{figure}
  \begin{center}
   \includegraphics[width=6cm]{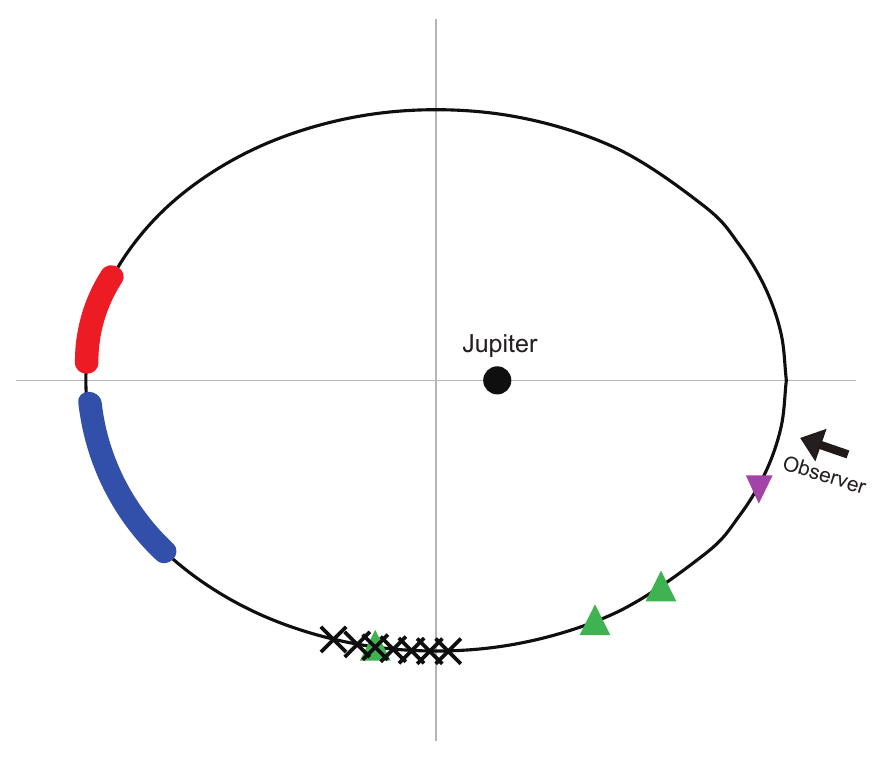}
  \end{center}
\caption{The true anomaly distribution for our observations (black cross), the auroral H and O results \citep{roth14b} (blue band), three measurements using transit observations \citep{sparks16} (green triangles), Keck/NIRSPEC observations \citep{paganini20} (red band) and JWST \citep{villanueva23} (purple inverted triangle). Black arrow indicates the direction of the line of sight during our observation.
 {Alt text: Map of Europa's true anomaly for observations on its elliptical orbit.} 
}
\label{fig:trueanomaly}
\end{figure}

We conducted high-dispersion spectroscopic observations in the \textit{L}-band using the Subaru Telescope/IRCS \citep{kobayashi00}. 
The observation date was July 17, 2021 (UT), when Europa was in a halfway from apojove to perijove (true anomaly $\sim$253\,–\,272$^{\circ}$, Table \ref{tab:observations}, Figure~\ref{fig:trueanomaly}), with Europa's heliocentric ($r_{\rm{h}}$) and geocentric distances ($\Delta$) at $r_{\rm{h}}$\,$\sim$\,5.03 and $\Delta$\,$\sim$\,4.17\,au, respectively, and a relative velocity to Earth of -29.6\,km s$^{-1}$.
We employed the cross-dispersed echelle spectroscopy mode with a 0.14"\,$\times$\,6.69" slit, providing spectral resolving power of $\lambda$/$\Delta\lambda$\,$\sim$\,20,000 in the \textit{L}-band. 
This setup allowed for detailed line-by-line and spatial analysis of water emissions to pinpoint a source location.

To enhance sensitivity by reducing additional background radiation from the sky and Europa's disk, we employed the AO188 combined with IRCS \citep{hayano10}. Our observations were carried out in the NGS mode, with Europa as the reference star for AO correction. The natural seeing ranged from $\sim$0.3"--0.9", and favorable weather conditions enabled effective AO performance throughout most of the observation period. This achieved spatial resolution close to the pixel sampling limit of 0.14", except during periods of low altitude for Europa early in the observation (Table \ref{tab:observations}).

\begin{figure}
  \begin{center}
   \includegraphics[width=8cm]{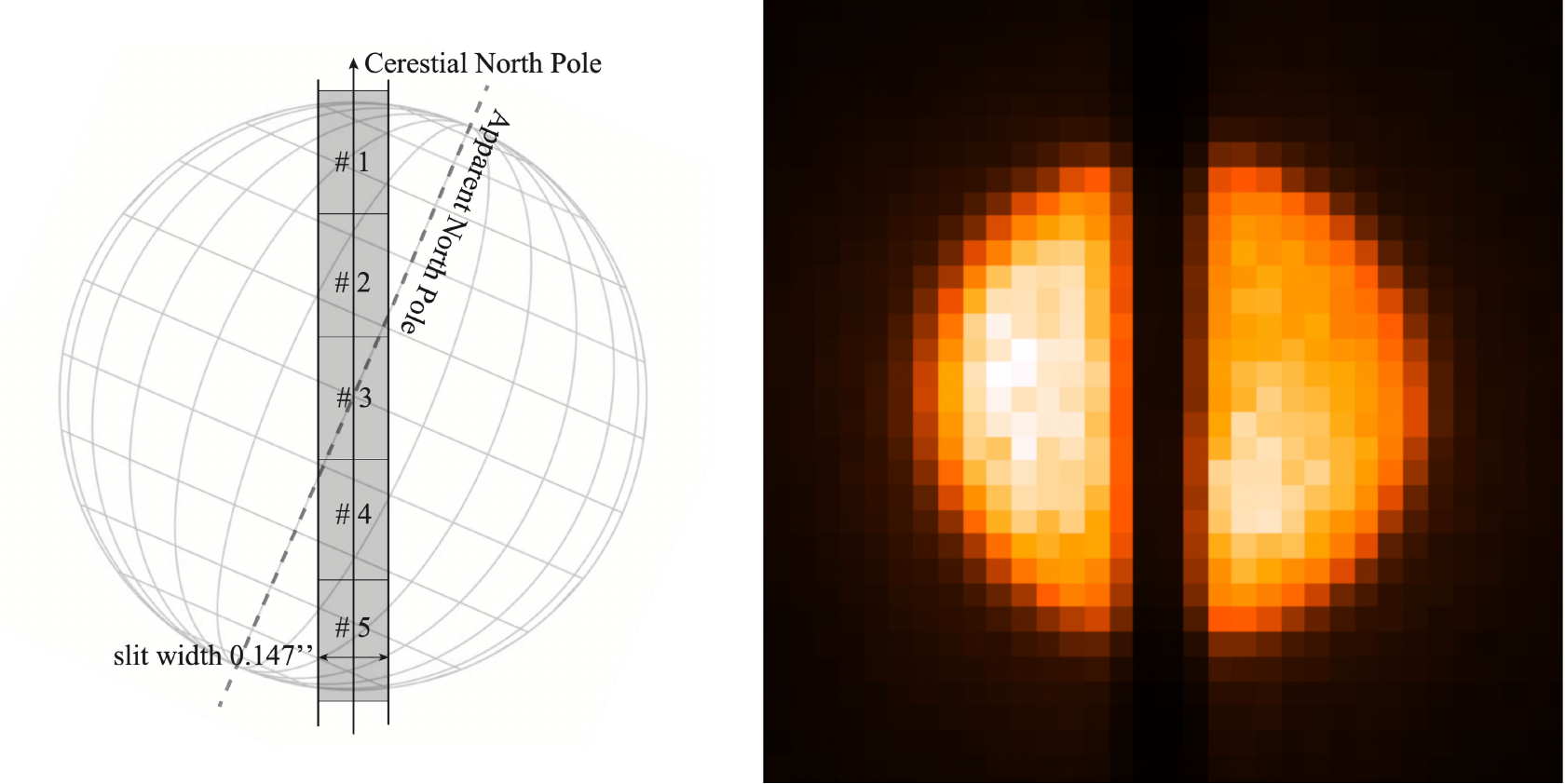}
  \end{center}
\caption{(left) Illustration of the alignment of the slit for our observations of Europa, and (right) an example of an actual slit viewer image during observations (captured at 12:20 UT). The apparent angular diameter was $\sim$1''032, and the apparent north pole position angle was 23.40$^\circ$ at the time of observations. We selected 0.14''$\times$6.69'' slit for our observations. Note that a K-band filter and an ND filter were employed in the slit viewer.
 {Alt text: Map (left)  and observational image (right).} 
}
\label{fig:alignment}
\end{figure}

\begin{figure}
  \begin{center}
   \includegraphics[width=8cm]{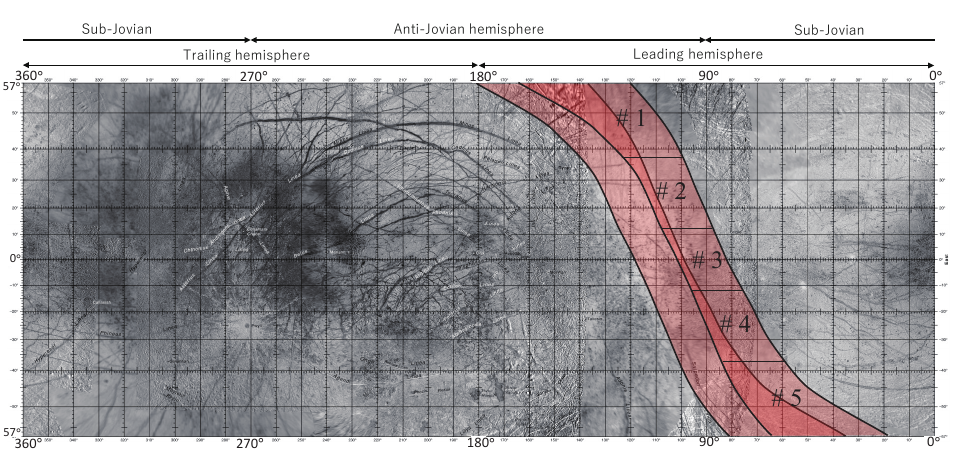}
  \end{center}
\caption{Observed region on Europa sampled by the slit of IRCS in red. The right and left hatched areas indicate the regions observed at 10:21 and 14:25, respectively. (Table \ref{tab:observations}).
 {Alt text: Map of Europa.} 
}
\label{fig:region}
\end{figure}

The slit position angle was aligned celestial North-South direction (PA=0\textdegree). The apparent north pole position angle was $\sim$23.4 degrees to the direction of the celestial north pole at the observations based on the NASA JPL Horizons (Figure \ref{fig:alignment}). The observation period was from 10:21 to 14:41 (UT), during which the observation area progressively shifted westward over time, and the center of the slit moved from $\sim$90 to $\sim$109 west longitude. As a result, the observed area of Europa scanned during our observations is depicted in Figure \ref{fig:region} as a red-shaded region.

Each exposure was set with an integration time of 120 seconds to ensure stable background conditions. This duration enabled precise subtraction of background emissions and mitigating saturation from sky emission lines. A-B-B-A dithering, a standard technique in near-infrared spectroscopy, was employed for spectral acquisition. The separation between A- and B-position was 3.6", which is adequate given Europa's apparent diameter. The total integration time was 92 minutes on source (see Table \ref{tab:observations}). 

In typical infrared spectroscopic observations, telluric standard stars are acquired to correct for atmospheric absorption lines in the spectra. We observed HD145964 (B9V) at different airmasses (1.4/1.3 at the start/end of observations) as a telluric and a photometric standard star. However, as described in the following section, synthetic spectra obtained by combining data from different regions on Europa provided more precise telluric corrections. Therefore, the standard star was used only for Europa's flux calibration.

%%%%%%%%%%%%%%%%%%%%%%%%%%%%%%%%%%%%%%%%%%%%%%%%%%%%%%%%%%%%%%
\subsection{Data reductions and analysis}\label{section2.2}
We conducted standard reduction procedures on the acquired spectral images, which included sky-background subtraction, flat-field correction, removal of pixels affected by cosmic rays and bad pixels, wavelength calibration, spatial and spectral alignment, and registration of individual A and B beams. This process resulted in obtaining two-dimensional spectra.

Subsequently, we co-added 3 or 4 ABBA spectra (each with integration times of 12 or 16 minutes) acquired during close observational periods. The slit center longitude of Europa has varied only 1 or 2 degrees across the combined dataset. The combined spectra were divided into five regions in the spatial direction, from which we extracted one-dimensional spectra. Each area was summed over 5 pixels in the spatial direction (0.267", $\sim$779\,km on Europa) and 2 pixels in the spectral direction (0.14", 423\,km on Europa). We note that the extracted spectra covered a 0.15" wider of Europa's disk ($\sim$1.0") to search for water in the upper layers of the plumes as well. 

We also determined a median spectrum of all extracted spectra to correct for atmospheric transmittance. As mentioned in section \ref{section2.1}, this correction was based on spectra acquired at almost the same observing time and airmass, allowing for a more reliable correction than standard stars. Europa's continuum component was subtracted using a third-order spline function fitting. The obtained spectra were flux-calibrated by comparing the spectra of standard stars (HD145964). We estimated the slit loss of the standard star by fitting a Gaussian profile to the observed PSF. The slit-loss correction was simultaneously applied during the flux calibration process. Finally, the flux-calibrated spectra were Doppler-shifted by the topocentric velocity of the Europa at the observations.

Note that our data reduction procedure is well-established and has been applied to multiple solar system objects to analyze water emissions \citep{Kawakita09}, as well as previous observations of Europa in the near-infrared \citep{paganini20}. Additional details of our data-reduction methods are given elsewhere (e.g., \cite{kobayashi10}). 

%%%%%%%%%%%%%%%%%%%%%%%%%%%%%%%%%%%%%%%
\setlength{\tabcolsep}{2pt}
\begin{longtable}{*{8}{c}}
\caption{Summaries of our observations.}\label{tab:observations} \\
\hline
Observation date & Integration & Relative & Wind & Mean seeing & Airmass & Sub-observer & True \\
and time\footnotemark[$*$] & time & humidity & (mph)\footnotemark[$\dag$] & (DIMM)\footnotemark[$\dag$] &  & longitude & anomaly \\ 
(UT) (HST)\footnotemark[$*$] & (seconds) & (\%)\footnotemark[$\dag$] &  &  &  & (degW) & (deg) \\
\endhead
\hline
\endfoot
\hline
\multicolumn{8}{l}{\footnotemark[$*$] These values represent the startpoint of data acquisition.}\\
\multicolumn{8}{l}{\footnotemark[$\dag$] These values obtained from Maunakea Weather Center (MKWC) (http://mkwc.ifa.hawaii.edu/current/seeing/index.cgi).}\\ 
\hline
\endlastfoot
\hline
17 July 2021 10:21 (00:21) & 720 & 10 & 3 & 0.38 & 1.54 & 90 & 253 \\
17 July 2021 11:02 (01:02) & 960 &  7 & 3 & 0.49 & 1.34 & 93 & 257 \\
17 July 2021 11:41 (01:41) & 960 & 12 & 3 & 0.74 & 1.24 & 96 & 260 \\
17 July 2021 12:20 (02:20) & 960 & 15 & 2 & 0.78 & 1.19 & 99 & 263 \\
17 July 2021 13:05 (03:05) & 960 & 24 & 1 & 0.83 & 1.18 & 102 & 266 \\
17 July 2021 13:45 (03:45) & 960 & 28 & 3 & 0.90 & 1.23 & 105 & 269 \\
17 July 2021 14:25 (04:25) & 960 & 29 & 3 & 0.61 & 1.29 & 108 & 272 \\
\hline
\end{longtable}
%%%%%%%%%%%%%%%%%%%%%%%%%%%%%%%%%%%%%%%%%%%%%%%%%%%%%%%%%%%%%%%%%%%%%%%%%%%%%%%%%%
%%%%%%%%%%%%%%%%%%%%%%%%%%%%%%%%%%%%%%%%%%%%%%%%%%%%%%%%%%%%%%%%%%%%%%%%%%%%%%%%%%%%%%%

%%%%%%%%%%%%%%%%%%%%%%%%%%%%%%%%%%%%%%%%%%%%%%%%%%%%%%%%%%%%%%%%%%%%%%%%%%%%%%%%%%%%%%%

%%%%%%%%%%%%%%%%%%%%%%%%%%%%%%%%%%%%%%%%%%%%%%%%%%%%%%%%%%%%%%%%%%%%%%%%%%%%%%%
%%%%%%%%%%%%%%%%%%%%%%%%%%%%%%%%%%%%%%%%%%%%%%%%%%%%%%%%%%%%%%%%%%%%%%%%%%%%%%%
\section{Results and Discussions}
Direct detection of water emissions from ground-based facilities is challenging due to severe water absorption. However, weak non-resonance fluorescence bands (hot-bands) induced by solar radiation can be observable even from the ground because water molecules in the atmosphere cannot absorb the energy emitted from such high vibrational energy states. 
Since there are continuous telluric absorption lines throughout the entire L-band, atmospheric conditions impact high-dispersion spectroscopy more significantly than instrumental effects, such as read noise and dark current. There are two main negative effects of atmospheric absorption on high-dispersion spectroscopy. First, stronger telluric absorption increases the thermal background, leading to reduced sensitivity. Additionally, the continuous telluric absorption makes it harder to determine the continuum level of the signal from the Europa disk, introducing systematic effect in detecting of absorption and emission lines.
Therefore, hot-band water emission lines are observed at wavelengths free from severe telluric absorption.
These hot-band lines exhibit prominent lines near 2.9\,$\mu$m, previously detected in Europa \citep{paganini20}. 

\begin{figure}
  \begin{center}
   \includegraphics[width=8cm]{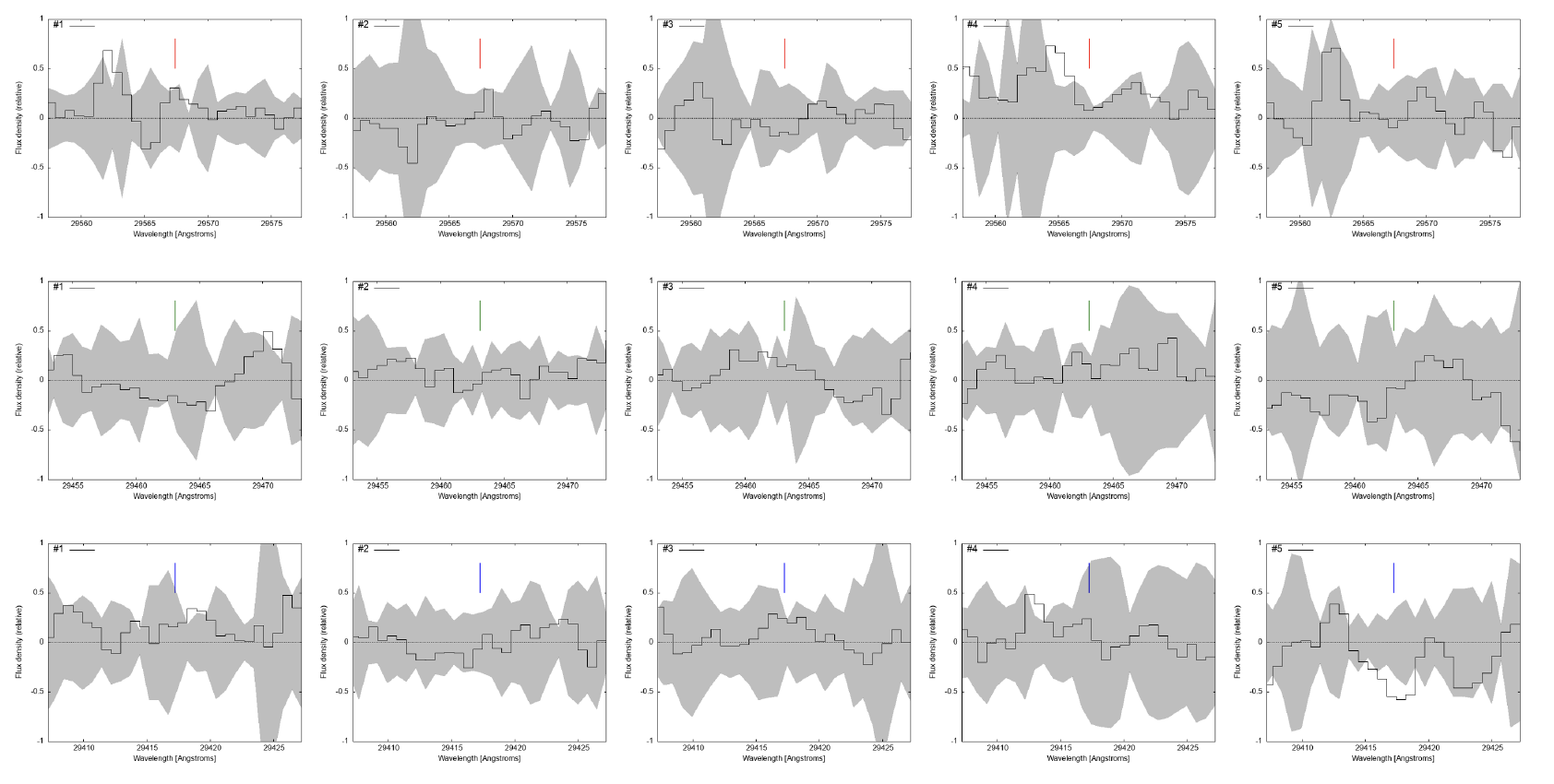}
  \end{center}
\caption{Examples of an integrated spectra and correlation analysis of water hot-band lines resulting from observations of Europa on 12:20 (UT) 17 July 2021. 
The red, green, and blue lines correspond to the expected wavelengths of 29567.4, 29463.1, and 29417.2 Å for Europa's hot-band lines, respectively.
The grey shaded area shows $\pm$3$\sigma$ uncertainties.
1$\sigma$ error levels are determined based on the variance at the same wavelength within 3 or 4 co-added spectra of close observational periods.
 {Alt text: Integrated spectrum (relative flux densities for each wavelength).} 
}
\label{fig:3lines}
\end{figure}

In this study, we investigated the spectra of three specific hot-band lines; (200)1$_{01}$-(001)2$_{02}$ at 29417.2 \AA, (200)2$_{21}$-(100)3$_{30}$ at 29463.1 \AA, and (200)2$_{12}$-(001)3$_{13}$ at 29567.4 \AA, respectively.
These lines exhibit high atmospheric transmittance ($\tau \sim$0.9) and significant fluorescence efficiencies (g-factor, see below), approximately (6--8)$\times$10$^{-9}$ photons s$^{-1}$ at $\sim$5 AU from the Sun.
Notably, these lines were previously studied by \citet{paganini20} including the emission line of (200)1$_{01}$-(001)2$_{02}$ at 29417.2 \AA\,being specifically detected. 
Thus, we can compare the variations in intensity and physical properties for these lines. 
The selected spectra of these three lines are shown in Figure \ref{fig:3lines}. 
As shown in this figure, no significant hot-band emission lines were detected either in each divided slit or the entire area covered by the slit during our observations. Hence, we determined the 1$\sigma$ error levels of flux based on the variance at three lines of wavelengths within three or four co-added spectra from close observation periods.

We derived 3$\sigma$ upper limits of the flux based on the median spectra of each region. 
The 3$\sigma$ upper limits of column density, $N_{\rm{col}}$ (m$^{2}$), total molecules, and production rates, $Q$ (molecules s$^{-1}$) using the flux of each line, $F_{\rm{lines}}$ (W\,m$^{-2}$). Column density can be derived using a well-established formalism;
\begin{equation}
N_{\rm{col}} = \frac{4 \pi \Delta^{2} F_{\rm{lines}}}{g_{\rm{lines}} A_{\rm{FOV}}},
\end{equation}
where the geocentric distance $\Delta$\,(m) represents the distance of Europa from Earth, while $A_{\rm{FOV}}$\,(m$^2$) denotes the area covered by our field of view. 
The fluorescence efficiencies (g-factor) of each line, $g_{\rm{lines}}$\,(W molecule$^{-1}$), is calculated based on a fluorescence excitation model (e.g., \cite{villanueva12}). In the models, it is generally assumed that the population of rotational states within the ground vibrational states follows a Boltzmann distribution at a rotational temperature ($T_{\rm{rot}}$) and is excited by solar UV. Thus, the model includes several parameters such as the distance from the Sun and $T_{\rm{rot}}$). 
For direct comparison with the previous observation that detected the prominent hot-band lines near 2.9\,$\mu$m \citep{paganini20}, we used a g-factor of 50\,K provided by that observation group (\cite{villanueva12}, $r_{\rm{h}}$ at 1au) with the corrections of heliocentric distance at the observations.%, https://astrobiology.gsfc.nasa.gov/Villanueva/spec.html. 
We also adopted 150\,K as the rotational temperature with our original water emission models \citep{Kawakita09} since the hot-band water would reflect the temperature of the plume.
Temperature of 50\,K is based on physical arguments considering a high-speed plume starting at 230\,K in the vent, then being cooled by the nozzle effect to $\sim$\,50\,K as it erupts out into space, dependent on the vent properties \citep{Berg16,DaytonOxland23}. 
In addition, this temperature is observationally expected in expanding gases in cometary atmospheres \citep{paganini20}.
In case of a weak nozzle effect in Europa's vent, the water plume temperature could be around 150 K \citep{Teolis17} as a higher end-member.

\begin{figure}
  \begin{center}
   \includegraphics[width=7cm]{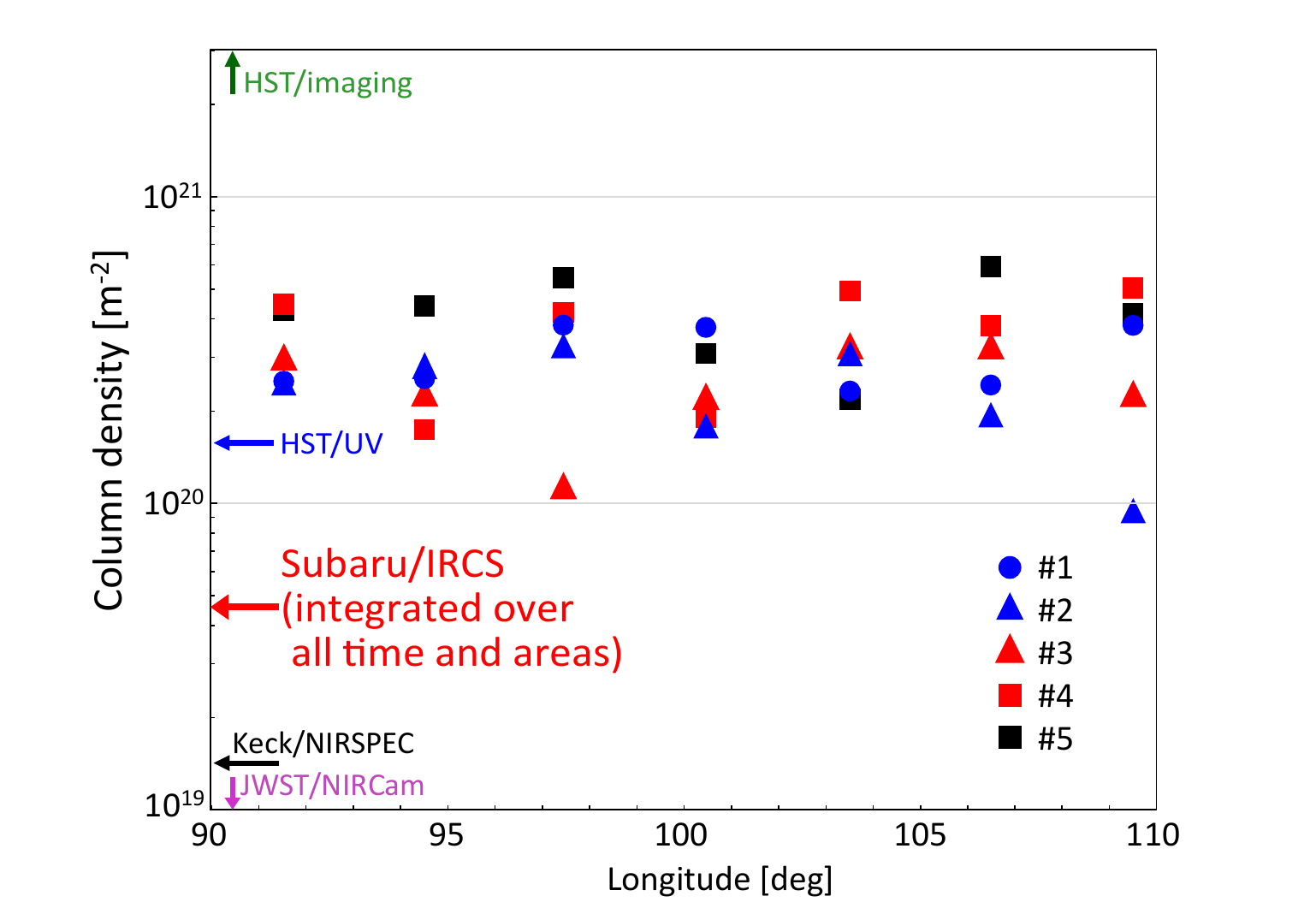}
  \end{center}
\caption{Estimated upper limit of the water molecules for each slit area, \#1$\sim$\#5, and for all pixels in the slit (red arrow) during our observations with other estimates by previous observations. The rotational temperature of the water molecules is assumed to be 50\,K. 
 {Alt text: Estimated upper limits for column density for Europa's longitudes.} 
}
\label{fig:columndensity}
\end{figure}

The total number of molecules is also derived by multiplying $N_{\rm{col}}$ by $A_{\rm{FOV}}$. Production rates, $Q$ (molecules s$^{-1}$), are also calculated to assess the gaseous production activity from celestial objects (e.g., comets), determined by the following formula;
\begin{equation}
Q = \frac{N_{\rm{col}} A_{\rm{FOV}}}{t},
\end{equation}
%where t is the lifetime of water molecules in the plume (887 sec, \cite{paganini20}). We statistically combined line-by-line analysis of three specific water emission lines and summarized 3$\sigma$ upper limits of column density, total molecules, and water production rates at T$_{rot}\,=\,$50\,K and 150\,K in (Table \ref{tab:results}). 
Here we focus on Europa's column density to compare it with previous studies.
Figure \ref{fig:columndensity} shows the upper limit of the column density of the water molecules that we estimated in each slit area, 9.46$\times$10$^{19}$ -- 5.92$\times$10$^{20}$\,m$^{-2}$ (corresponding to 3.12$\times$10$^{31}$ -- 1.95$\times$10$^{32}$ molecules).
The column density estimated in the entire area covered by the slit is 4.61$\times$10$^{19}$\,m$^{-2}$ (total molecules, 8.11$\times$10$^{31}$).

\begin{figure}
  \begin{center}
   \includegraphics[width=7cm]{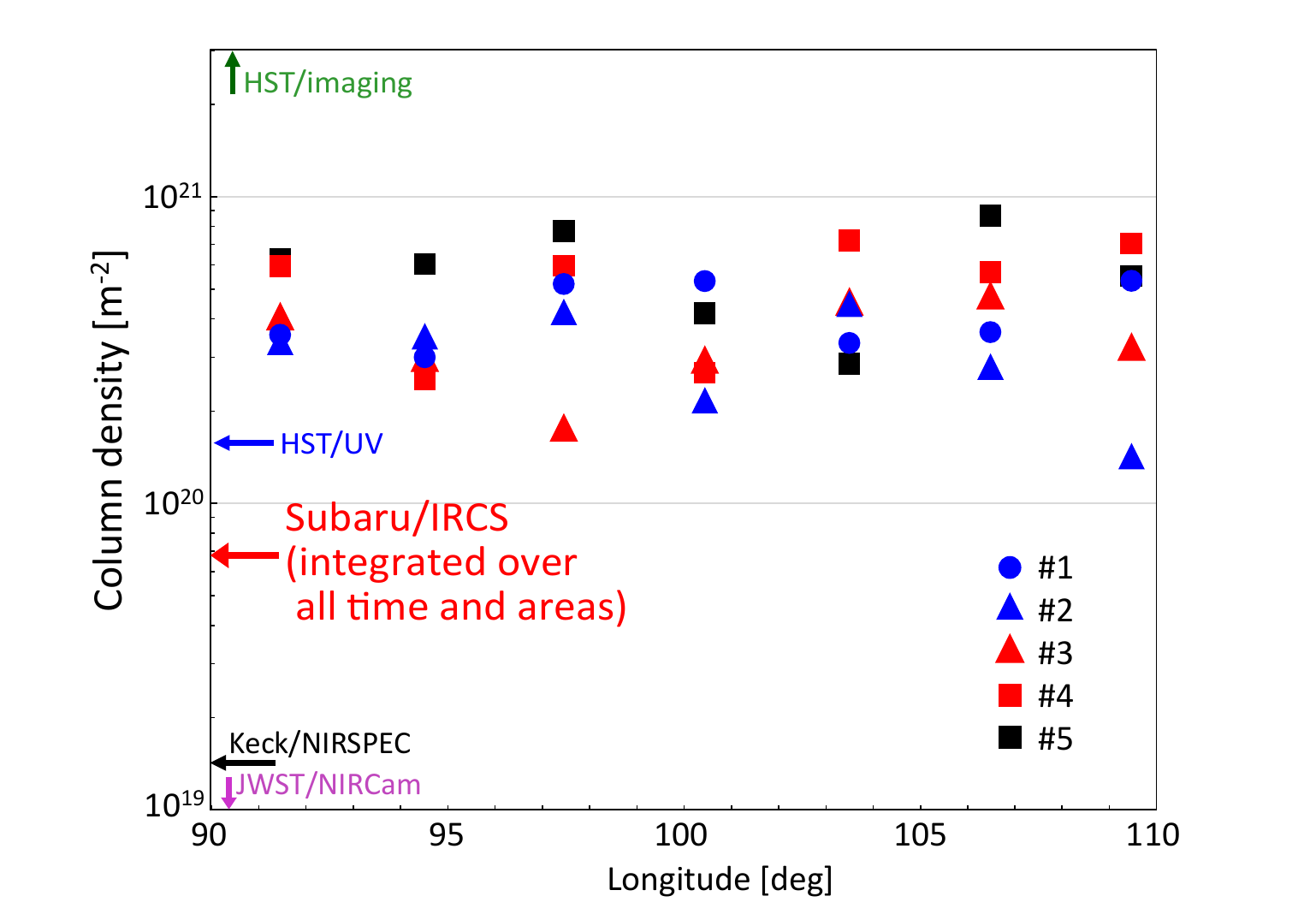}
  \end{center}
\caption{Same as Fig. \ref{fig:columndensity} but with rotational temperature of the water molecule of 150\,K. Column density estimated for all pixels is 6.90$\times$10$^{19}$\,m$^{-2}$.
 {Alt text: Estimated upper limits for column density for Europa's longitudes.} 
}
\label{fig:columndensity150K}
\end{figure}

In the case of T$_{\rm{rot}}$\,=\,150\,K, the upper limits of the water molecule are slightly larger than that of T$_{\rm{rot}}$\,=\,50\,K, (1.49 -- 8.69)$\times$10$^{20}$\,m$^{-2}$ (Table \ref{tab:results}, Figure \ref{fig:columndensity150K}). 
The column density estimated in the entire area covered by the slit is 6.90$\times$10$^{19}$\,m$^{-2}$. 
The discrepancy in these values between different rotational temperatures arises from differences in the g-factor. The emission lines focused on in this study predominantly originate from lower rotational states within the ground vibrational levels due to selection rules. Hence, higher rotational temperatures correspond to reduced fluorescence efficiency. The g-factor at T$_{\rm{rot}}$\,=\,50\,K used in this study exhibits a roughly 1.5 times difference compared to that at 150\,K.
Assuming a rotational temperature of 150 K, fluorescence efficiency increases significantly for transitions involving higher energy levels compared to the T$_{\rm{rot}}$\,=\,50\,K. We also searched for emission lines with the $g_{\rm{lines}} > 1\times 10^{-9}$ at 5 au; however, no emission lines were detected.

The performance of high-dispersion spectroscopic observations in the L band from the ground is significantly influenced by various observational conditions, such as humidity, airmass, and the extent of cloud cover.
Additionally, although both the Subaru/IRCS and Keck/NIRSPEC observations employ a similar approach to detect fluorescent emissions from water vapor using high-dispersion spectroscopy, only the Subaru observations spatially resolve the Europa disk.
Therefore, we derived the column density of the water vapor instead of the total water molecule count in order to compare the Subaru/IRCS observations with previous studies.
Considering the fact that the integration time of our observation is almost the same as that of the Keck observation, the Subaru telescope exhibits approximately one-third the sensitivity of Keck.
Assuming a similar amount of water molecules suggested by the Keck observations was present during our observations, the sensitivity achieved in this study would not be sufficient to detect them at those level.

\begin{figure}
  \begin{center}
   \includegraphics[width=8cm]{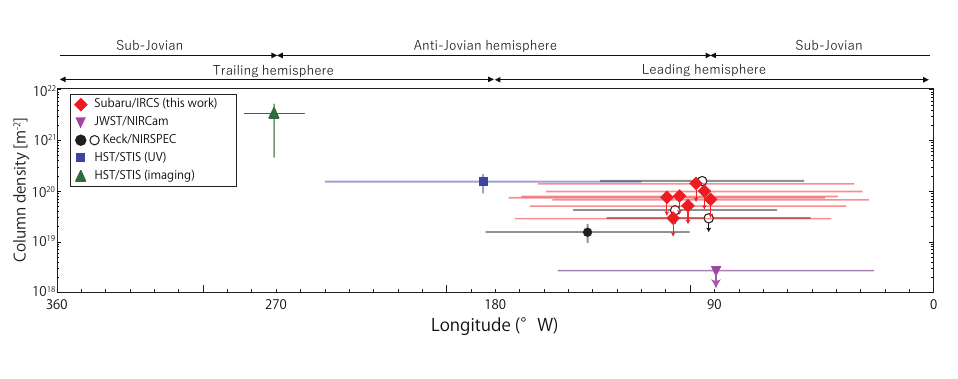}
  \end{center}
\caption{Estimated upper limits for column density of water molecules (red diamonds, T$_{rot}$\,=\,50\,K) versus longitudinal coverage with previous estimates by HST/STIS UV (green triangle) \citep{roth14b}, HST/STIS imaging (blue square) \citep{sparks16}, KECK/NIRSPEC (detected value as black filled-circle and upper limits as unfilled circles) \citep{paganini20} and JWST/NIRCam observations (purple inverted triangle) \citep{villanueva23}.
 {Alt text: Map for upper limits for column density.} 
}
\label{fig:longitudinalcoverage}
\end{figure}

Figure \ref{fig:longitudinalcoverage} shows the upper limits of the column density in the entire area covered by slit for each 7 integration, with longitudinal coverage on Europa.
Our estimates for the column density, 2.98$\times$10$^{19}$ -- 1.40$\times$10$^{20}$\,m$^{-2}$, is smaller than that by UV observation using the HST \citep{roth14b}, indicating that the water vapor was not detected at the levels inferred from the HST data.
These estimated upper limits are a factor of 11 and 52 larger than the previous JWST observations of the leading hemisphere of 2.71$\times$10$^{18}$\,m$^{-2}$ \citep{villanueva23}.
Our observed area started from the center of the leading hemisphere, which is a similar longitude as the JWST observation \citep{villanueva23}, and shifted westward to the anti-Jovian hemisphere where previous water detections by HST/UV \citep{roth14b} and Keck/NIRSPEC \citep{paganini20} as time proceeded. 
Although our observational longitudes are partially overlapped to the area where water molecules were detected by Keck/NIRSPEC, our observed area in the southern hemisphere is east of the central longitude, away from the Keck observation area.
Thus, even if the plume was occurring in the region seen by Keck, it might have been outside of our field of view.
\citet{paganini20} reported that three observations by Keck/NIRSPEC (26 April 2016, 15 and 19 March 2017) at similar longitudes to our observations were undetected, and suggested no correlation between plume activity and source location, indicating that Europa is unlikely to have large active fractures that could serve as fixed plume source, in contrast to Enceladus's "tiger stripes". 
Europa's plume material might be sporadically erupted due to tidal stress from smaller-scale sources than the resolution of existing surface image.
At the time of our observation, Europa was far from the apojove, where signals were detected by HST and Keck, and was halfway between perijove and apojove.
It may imply that surface tensile stress is insufficient at this orbital phase.

The JWST conducts space-based observations focusing on the fundamental vibrational band, which complicates direct comparisons with ground-based observations. In contrast, observations with the Keck/NIRSPEC target the hot-band, involving seven distinct lines. Within our current observational settings, only three of these lines are covered; however, extending coverage to all seven lines would enhance the S/N ratio by a factor of 1.5. To achieve a sensitivity comparable to that of the Keck/NIRSPEC, based solely on the number of emission lines observed, would require more than 30 lines. The availability of hot-band lines exhibiting a significant g-factor, $\sim$10$^{-9}$ at 5\,AU, is limited. Moreover, if there is temporal variability in the water, synchronic observations become crucial, necessitating innovate approaches to equipment and settings that can cover a broad wavelength range simultaneously.

%%%%%%%%%%%%%%%%%%%%%%%%%%%%%%%%%%%%%%%
\begin{longtable}{*{12}{c}}
\caption{3$\sigma$ Upper limits of the water column density, production rates, and total molecules of Europa on 17 July 2021 (UT). Slit areas of $\#$1 - $\#$5 are same as in Figures \ref{fig:alignment} and \ref{fig:region}.}
\label{tab:results}
\hline
\endhead
\hline
\endfoot
\hline
\endlastfoot
& \multicolumn{11}{c}{Column density [$\times$10$^{20}$ m$^{-2}$]} \\
Obs. time & \multicolumn{5}{c}{$T_{rot} =$ 50K} & & \multicolumn{5}{c}{$T_{rot} =$ 150K} \\
\cline{2-6}\cline{8-12}
[UT] & $\#$1 & $\#$2 & $\#$3 & $\#$4 & $\#$5 & & $\#$1 & $\#$2 & $\#$3 & $\#$4 & $\#$5 \\
\hline
10:21 & $<$2.50 & $<$2.47 & $<$3.00 & $<$4.45 & $<$4.26 & & $<$3.62 & $<$3.44 & $<$4.16 & $<$6.00 & $<$6.31 \\
11:02 & $<$2.55 & $<$2.83 & $<$2.29 & $<$1.74 & $<$4.40 & & $<$3.07 & $<$3.59 & $<$3.06 & $<$2.60 & $<$6.09 \\
11:41 & $<$3.81 & $<$3.26 & $<$1.14 & $<$4.20 & $<$5.44 & & $<$5.26 & $<$4.28 & $<$1.83 & $<$6.02 & $<$7.77 \\
12:20 & $<$3.74 & $<$1.79 & $<$2.23 & $<$1.90 & $<$3.08 & & $<$5.38 & $<$2.24 & $<$3.02 & $<$2.74 & $<$4.24 \\
13:05 & $<$2.32 & $<$3.08 & $<$3.26 & $<$4.92 & $<$2.18 & & $<$3.41 & $<$4.56 & $<$4.64 & $<$7.25 & $<$2.92 \\
13:45 & $<$2.43 & $<$1.94 & $<$3.27 & $<$3.79 & $<$5.92 & & $<$3.69 & $<$2.86 & $<$4.85 & $<$5.74 & $<$8.69 \\
14:25 & $<$3.80 & $<$0.95 & $<$2.28 & $<$5.03 & $<$4.16 & & $<$5.38 & $<$1.49 & $<$3.32 & $<$7.09 & $<$5.58 \\
& \multicolumn{11}{c}{Production rates [$\times$10$^{29}$ molecules m$^{-1}$]} \\
Obs. time & \multicolumn{5}{c}{$T_{rot} =$ 50K} & & \multicolumn{5}{c}{$T_{rot} =$ 150K} \\
\cline{2-6}\cline{8-12}
[UT] & $\#$1 & $\#$2 & $\#$3 & $\#$4 & $\#$5 & & $\#$1 & $\#$2 & $\#$3 & $\#$4 & $\#$5 \\
\hline
10:21 & $<$0.93 & $<$0.92 & $<$1.12 & $<$1.66 & $<$1.59 & & $<$1.44 & $<$1.37 & $<$1.65 & $<$2.38 & $<$2.51 \\
11:02 & $<$0.95 & $<$1.05 & $<$0.85 & $<$0.65 & $<$1.64 & & $<$1.22 & $<$1.42 & $<$1.22 & $<$1.03 & $<$2.42 \\
11:41 & $<$1.42 & $<$1.22 & $<$0.43 & $<$1.56 & $<$2.03 & & $<$2.09 & $<$1.70 & $<$0.73 & $<$2.39 & $<$3.08 \\
12:20 & $<$1.39 & $<$0.67 & $<$0.83 & $<$0.71 & $<$1.15 & & $<$2.13 & $<$0.89 & $<$1.20 & $<$1.09 & $<$1.68 \\
13:05 & $<$0.86 & $<$1.15 & $<$1.21 & $<$1.83 & $<$0.81 & & $<$1.35 & $<$1.81 & $<$1.84 & $<$2.88 & $<$1.16 \\
13:45 & $<$0.90 & $<$0.72 & $<$1.22 & $<$1.41 & $<$2.20 & & $<$1.46 & $<$1.14 & $<$1.92 & $<$2.28 & $<$3.45 \\
14:25 & $<$1.41 & $<$0.35 & $<$0.85 & $<$1.87 & $<$1.55 & & $<$2.13 & $<$0.59 & $<$1.32 & $<$2.81 & $<$2.21 \\
 & \multicolumn{11}{c}{Total molecules [$\times$10$^{32}$ molecules]} \\
Obs. time & \multicolumn{5}{c}{$T_{rot} =$ 50K} & & \multicolumn{5}{c}{$T_{rot} =$ 150K} \\
\cline{2-6}\cline{8-12}
[UT] & $\#$1 & $\#$2 & $\#$3 & $\#$4 & $\#$5 & & $\#$1 & $\#$2 & $\#$3 & $\#$4 & $\#$5 \\
\hline
10:21 & $<$0.83 & $<$0.82 & $<$0.99 & $<$1.47 & $<$1.41 & & $<$1.28 & $<$1.21 & $<$1.46 & $<$2.11 & $<$2.22 \\
11:02 & $<$0.84 & $<$0.94 & $<$0.76 & $<$0.57 & $<$1.45 & & $<$1.08 & $<$1.26 & $<$1.08 & $<$0.92 & $<$2.15 \\
11:41 & $<$1.26 & $<$1.08 & $<$0.38 & $<$1.39 & $<$1.80 & & $<$1.85 & $<$1.51 & $<$0.65 & $<$2.12 & $<$2.74 \\
12:20 & $<$1.24 & $<$0.59 & $<$0.74 & $<$0.63 & $<$1.02 & & $<$1.89 & $<$0.79 & $<$1.06 & $<$0.96 & $<$1.49 \\
13:05 & $<$0.77 & $<$1.02 & $<$1.08 & $<$1.63 & $<$0.72 & & $<$1.20 & $<$1.60 & $<$1.63 & $<$2.55 & $<$1.03 \\
13:45 & $<$0.80 & $<$0.64 & $<$1.08 & $<$1.25 & $<$1.95 & & $<$1.30 & $<$1.01 & $<$1.71 & $<$2.02 & $<$3.06 \\
14:25 & $<$1.25 & $<$0.31 & $<$0.75 & $<$1.66 & $<$1.37 & & $<$1.89 & $<$0.52 & $<$1.17 & $<$2.49 & $<$1.96 \\
\hline
\end{longtable}

%%%%%%%%%%%%%%%%%%%%%%%%%%%%%%%%%%%%%%%%%%%%%%%%%%%%%%%%%%%%%%%%%%%%%%%%%%%%%%%%%%
%%%%%%%%%%%%%%%%%%%%%%%%%%%%%%%%%%%%%%%%%%%%%%%%%%%%%%%%%%%%%%%%%%%%%%%%%%%%%%%%%
%%%%%%%%%%%%%%%%%%%%%%%%%%%%%%%%%%%%%%%%%%%%%%%%%%%%%%%%%%%%%%%%%%%%%%%%%%%%%%%%%
\clearpage
%%%%%%%%%%%%%%%%%%%%%%%%%%%%%%%%%%%%%%%%%%%%%%%%%%%%%%%%%%%%%%%%%%%%%%%%%%%%%%%%%%%%%%%
%%%%%%%%%%%%%%%%%%%%%%%%%%%%%%%%%%%%%%%%%%%%%%%%%%%%%%%%%%%%%%%%%%%%%%%%%%
\section{Conclusions}
We conducted high-dispersion spectroscopic observations in the L-band using the Subaru Telescope/IRCS to directly search for water emission on Europa and to explore spatial variations in plume activity.
Our observations, utilizing the high spectral and spatial resolution, as well as the sensitivity of Subaru/IRCS, have enabled novel approaches for spatially resolved searches for water plumes on Europa.
Within our detection limits and time of observation, no evidence was found for the presence of water molecules. 
The derived upper limit for water vapor is 9.46$\times$10$^{19}$ -- 5.92$\times$10$^{20}$\,m$^{-2}$ in each divided slit area and 4.61$\times$10$^{19}$\,m$^{-2}$ in the entire area covered by the slit and at an assumed rotational temperature of 50K.

Identifying plume sources and understanding their extent, distribution, and temporal variability will advance our understanding of the still intriguing Jupiter's moon Europa and its potential for habitability.
Our Subaru/IRCS observations achieved an upper limit of water molecules that are more sensitive than previous estimates from HST/UV observations, while less sensitive by a factor of 3 than the Keck telescope and by one order or more than the James Webb Space Telescope (JWST) observations.

Our results are consistent with previous observations, confirming Subaru/IRCS as a viable tool for searching for water plumes on icy bodies. Further searches across different surface areas and orbital phases on Europa are encouraged to better identify potential plume sources and understand their extent, distribution, and variability.
\clearpage
%%%%%%%%%%%%%%%%%%%%%%%%%%%%%%%%%%%%%%%%%%%%%%%%%%%%%%%%%%%%%%%%%%%%%%%%%%
%\section{Appendix}
%To be added.
%%%%%%%%%%%%%%%%%%%%%%%%%%%%%%%%%%%%%%%%%%%%%%%%%%%%%%%%%%%%%%%%%%%%%%%%%%
\begin{ack}
We thank anonymous reviewer for constructive comments which greatly improved the manuscript.
This study was supported by KAKENHI from the Japan Society for Promotion of Science (Grant No. JP22K03700, JP17K05635 and JP22740285).
\end{ack}
%%%%%%%%%%%%%%%%%%%%%%%%%%%%%%%%%%%%%%%%%%%%%%%%%%%%%%%%%%%%%%%%%%%%%%%%%%
\bibliographystyle{apj}
\bibliography{jkbibfile}

\end{document}